# ACTOR GARBAGE COLLECTION IN DISTRIBUTED SYSTEMS USING GRAPH TRANSFORMATION

B. Seetha Lakshmi, C.D. Balapriya, R.Soniya

KLN College of Information Technology

Pottapalayam, Sivagangai District, Tamil Nadu, India

seethasee1976@rediffmail.com

## ABSTRACT

*A lot of research work has been done in the area of Garbage collection for both uniprocessor and distributed systems. Actors are associated with activity (thread) and hence usual garbage collection algorithms cannot be applied for them. Hence a separate algorithm should be used to collect them. If we transform the active reference graph into a graph which captures all the features of actors and looks like passive reference graph then any passive reference graph algorithm can be applied for it. But the cost of transformation and optimization are the core issues. An attempt has been made to walk through these issues.*

## 1. INTRODUCTION

When an object is no longer referenced by a program, the heap space it occupies can be recycled so that the space is made available for subsequent new objects. If there is no automatic storage reclamation then the programmer has to manually find the objects which are unused and has to collect them, which may be error prone and time consuming.

This paper is divided into three sections. In the first section the fundamentals of Actor System and Traditional Passive objects are being dealt with. Secondly, Transformation Algorithms are discussed. Third section thoroughly explains about the cost and optimization of transformation algorithm.

### 1.1 Distributed Garbage Collection

In the distributed systems can support both passive and active objects. Active objects relate to actors, and we use actors to refer to them. One major difference between actors and passive objects is the thread of control. A passive object is operated by external threads, which can create new objects, add new references, or delete references. If an object can be possibly manipulated by the external threads of control, it is live; otherwise it is garbage. On the other hand, an actor has an internal thread, it can be manipulated by the external thread s of control and at the same time it can also manipulate other objects provided if it is not in an blocked state. Hence an actor is live if it can be manipulated by external threads or if it can manipulate other object's thread otherwise it is garbage. Both active and passive objects can become garbage, and require a garbage collection mechanism to reclaim them.

### 1.1 Passive Object Reference Graph

Figure 1: Passive Reference Graph

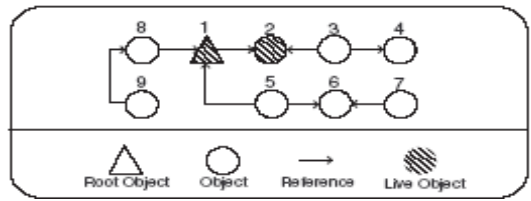

The above diagram is a passive object reference graph in which root objects are shown as triangles and others as circles. The objects (1,2) which can be reached from root objects were not garbage.

### 1.2 Actor Reference Graph

Figure 2: Active Reference Graph

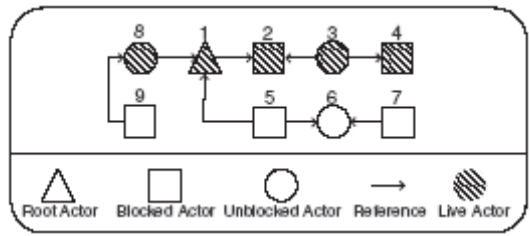

The above is an actor reference graph. Actors 3,4,8 are live because they can potentially sent messages to the root. For example, even thought he actor 3 cannot send message to the root directly, it can do so using object 2 (that is in 3 can activate 2 by calling it, which can be referred by root actor) indirectly.

### 1.3 Terminologies

The principle features and terminology of the actor model which relate to the garbage collection problem are these:

Passive Object: A passive object is one that only speaks when spoken to. i.e., Only responds and calls other functions on other objects, when one of its own functions is called. In essence, a traditional programming object

Active Object: An active object has a mind and life of its own. It owns its own thread of control, notionally associated with its own mini address space

Actor: A concurrently active object. There are no passive entities. Each actor is uniquely identified by the address of its single mail queue.

Acquaintance: Actor B is an acquaintance of actor A if B's mail queue address is known to actor A.

Inverse acquaintance: if actor A is an acquaintance of actor B, then actor B is an inverse acquaintance of A.

Acquaintance list: a set of mail queue addresses including any mail queue address contained in a message on the actors mail queue or in transit to the mail queue. This accounts for delays in message processing.

Blocked actor: All behaviors are blocked.

Active actor: an actor with at least one active behavior.

Root actors: An actor designated as being "always useful." Examples of root actors are those which have the ability to directly affect real-world through sensors, actuators, I/O devices, users, etc.

### 1.4 Need for Graph Transformation

When a system contains both active and passive object garbage then we have to use active object garbage collection algorithm to collect actor garbage and use passive object garbage collection algorithm to collect passive object garbage. Instead of using two algorithms we can use transformation algorithm to transform active object graph into passive object graph. In the next section let us discuss about two transformation algorithms. They are

       1. Transformation Algorithm by Vardhan and Agha.

       2. Transformation algorithm by Wei-Jen Wang et al.

## 2. TRANSFORMATION ALGORITHMS

### 2.1 Transformation Algorithm by Vardhan and Agha

The method proposed by Vardhan and Agha performs transformation of the actor reference graph which captures all the information necessary for actor GC, and makes it possible to apply a garbage collection algorithm for passive objects to the transformed graph in order to collect garbage actors. The transformation represents each actor in the original graph by a pair of nodes in the transformed graph. References between nodes in the transformed graph are derived using rules which depend not only on the actors to which know a particular actor, but also on which actors it knows; and whether or not that actor has messages pending in its mail queue.

**Rules for Transformation**

1. For every actor named a in original Actor graph, there are two corresponding nodes: Original Object and its Mail queue.

       Figure 3: Rule 1

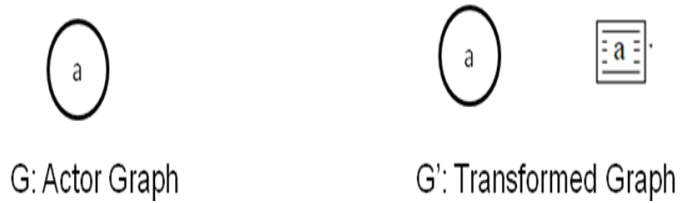

G: Actor Graph  G': Transformed Graph

2. For every root actor there is an equivalent object and its mail queue object.

Figure 4: Rule 2

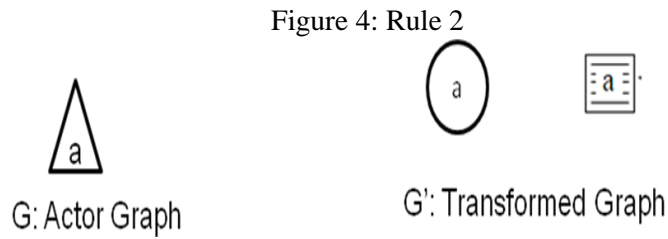

G: Actor Graph  G': Transformed Graph

3. If an actor a is unblocked, there is an edge from its mailqueue to itself in the transformed graph.

Figure 5: Rule 3

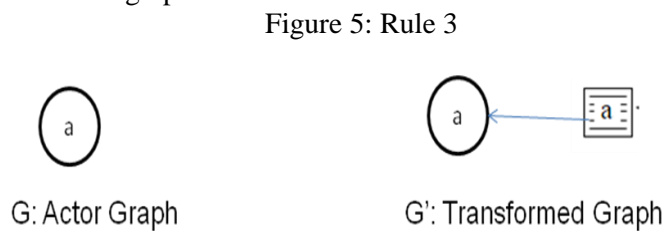

G: Actor Graph  G': Transformed Graph

4. If an actor a has a reference to an actor b, there is an edge from original object to its mailqueue and to the original object.

Figure 6: Rule 4

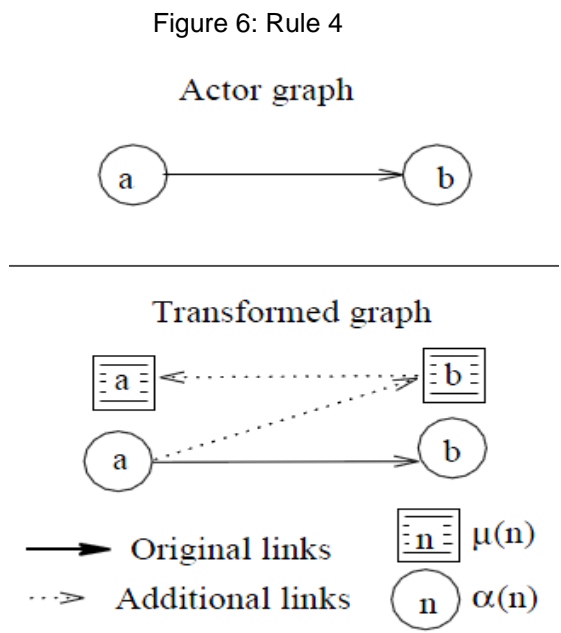

**Example**

In the below figure for actor names i = {1, 6, 10, 12} which are unblocked, there is an edge from µ(i) to α(i). Looking at this graph we can see that a garbage collector for passive objects would regard α(1), α(2), α(3), α(4), α(5), α(6) and α(8) as live and all others objects in A' as garbage. A look at the original actor-reference graph shows that it is exactly actors 1, 2, 3, 4, 5, 6 and 8 that are live. Of special interest is α(6) in the transformed graph. Because α(6) has a reference from µ(6) which is reachable from µ(1) (the root), it is correctly identified as being live. The reader can also note that, although µ(7) is reachable in the transformed graph, α(7) is not. By step 4 of Algorithm 1, it is α(7) that is used for deciding garbage status of actor 7 and hence 7 is correctly identified as garbage.

Figure 7: Example

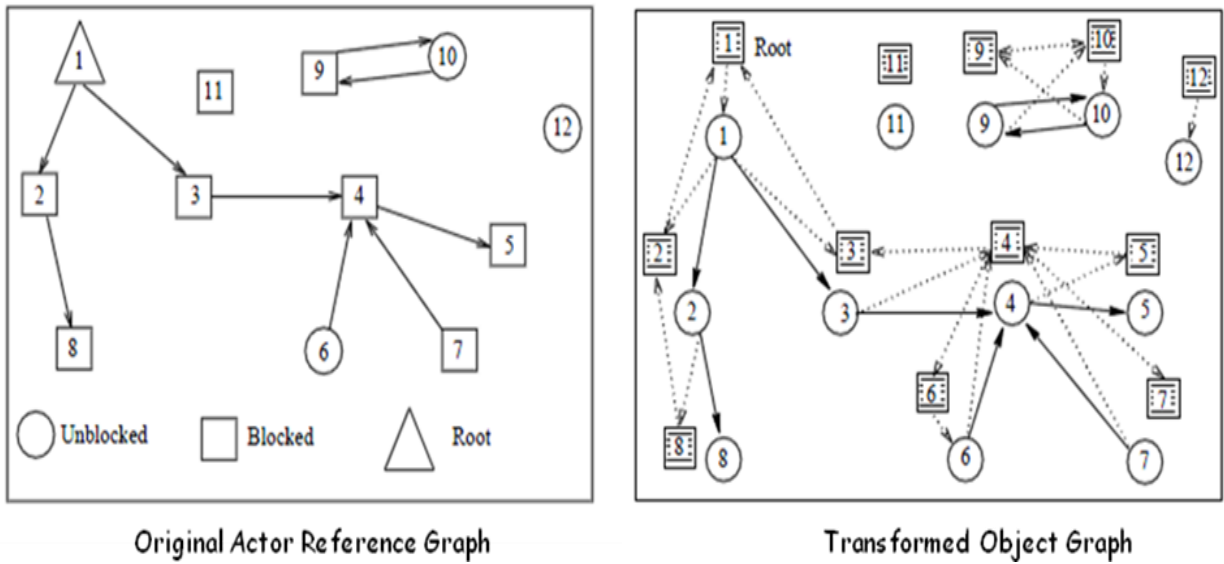

## 2.2 Transformation Algorithm by Wei Jen Wang

The essential concept of passive object garbage lies in the idea of the possibility of object manipulation. Objects that can be manipulated by the thread of control of the application are live; otherwise they are garbage. Root objects are those which can be directly accessed by the thread of control, while transitively live objects are those transitively reachable from the root objects by following references. The problem of passive object garbage collection and active garbage collection can be represented as a graph problem. Hence if we transform active reference graph into passive reference graph then we can apply any one of the passive garbage collection algorithm to collect garbage.

### 2.2.1 Transformation by Direct Back Pointers to Unblocked Actors.

This is a much easier approach to transform actor garbage collection into passive object garbage collection, by making

$$E' = E \cup \overline{\{a_q a_u} \mid a_u \in (U \cup R) \wedge a_u \rightsquigarrow a_q\}.$$

Figure 8: Example

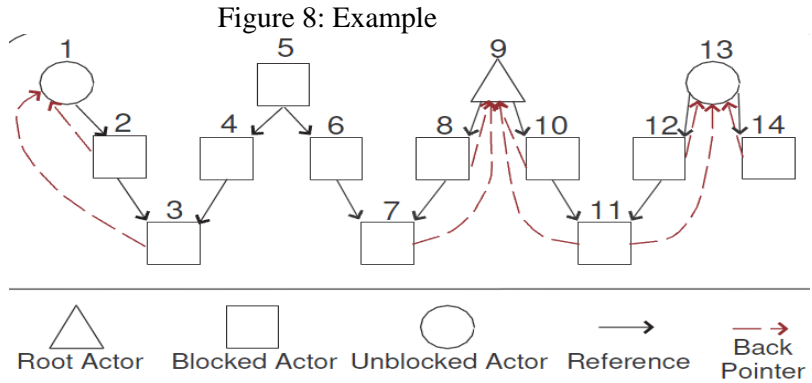

For example, in the above Fig.5, Actors 2 and 3 have back pointers to Unblocked Actor 1 because they are reachable from Actor 1. Actor 11 has a back pointer to Root Actor 9 and another one to Unblocked Actor 13 for the same reason. Actor 3 does not have a back pointer to Actor 5 because Actor 5 is neither a root nor an unblocked actor. Notice the use of term back pointers to describe the newly added references is to avoid ambiguity with the term in-verse references.

### 2.2.2 Transformation by Indirect Back Pointers to Unblocked Actors.

This is an another similar approach to transform actor garbage collection into passive object garbage collection,

$$E' = E \cup \overline{\{a_q a_p\} \mid a_u \in (U \cup R) \wedge \overline{a_p a_q} \in E \wedge a_u \rightsquigarrow a_p\}}.$$

Figure 9: Example

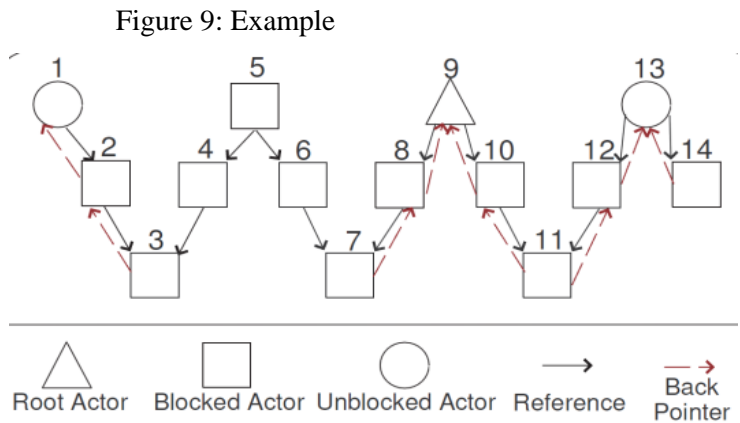

For example, in the above Fig.6, Actor 2 has back pointers to Unblocked Actor 1 and Actor 3 has back pointers to Actor 2 because they are reachable from Actor 1. The newly added back pointers will create a corresponding counter-directional path of a path from an unblocked/root actor to another actor which is reachable from the unblocked/root actor. Similarly, Actor 11 has a new counter-directional path to Root Actor 9 and another one to Unblocked Actor 13.

# 3. COST AND OPTIMIZATION

## 3.1 Transformation by Abhay Vardhan *et. al.*

### 3.1.1 Analysis of Cost

To perform analysis of cost of GC (Garbage Collection) an actor program wa run on the Actor Foundry. The program implements an exhaustive search solution for the 5-queens and 4-queens problem. The problem is to put 5 queens on an 5 by 5(4 queens on an 4 by 4) chess-board such that no queen is under attack from another one according to the rules of chess. In the implementation a single actor, C, starts the computation with an empty chess board. It places a single queen in one of the squares on the first row and creates an actor to solve the remainder of the problem. One actor is created for every square on the first row. When a newly created actor receives a partially filled chess-board it places queens on the row following the rows that have already been filled and spawns additional actors to do the remainder of the computation. If an actor manages to fill all rows, it sends a message to C notifying it of the solution. The program generated a large amount of garbage.

Table 1 : Timings for 5-Queens Problem(Single Host)

| No. | Total time without GC | Total time with GC | Ratio |
|---|---|---|---|
| 1 | 1.40 seconds | 2.235 seconds | 1.596 |

Table 1 shows the Timings for 5 queens problem on a single host. Table 2 shows the break cost of GC for 5 queens problem on a single host. Table 3 shows the Timings for 4 queens problem on a network which contains 2 host.

Table 2 : Breakup cost for 5-Queens Problem(Single Host)

| No. | Explanation | Total time | Overhead ratio |
|---|---|---|---|
| 1 | All parts of GC running | 2.235 seconds | 1.59 |
| 2 | No marking phase during execution | 2.160 seconds | 1.54 |
| 3 | No acquaintances are saved | 1.965 seconds | 1.40 |
| 4 | Loss of Actor references not detected | 1.395 seconds | 0.99 |

Table 3 : Timings for 4-Queens Problem(Two Hosts)

| No. | Total time without GC | Total time with GC | Overhead ratio |
|---|---|---|---|
| 1 | 2.834 seconds | 3.741 seconds | 1.32 |

Experimental results indicate that the ratio of time taken with GC running and without GC is seen to be 1.6 for a single host case and 1.3 for the case of a network with two hosts.

### 3.1.2 Analysis of Performance

Table 4 shows the various parameters which affect the performance together with the advantages, disadvantages and a possible suggesstion to overcome them.

Table: 4 Various Performance Parameters

| Parameter | Advantage | Disadvantage | Suggession |
|---|---|---|---|
| Mail Queue Objects | To know the blocked or unblocked status of the objects | Extra space occupied by mail queue object is an Overhead | Reference between the mailqueue objects and reference between the can be maintained as a bit in the original object itself. |
| Inverse Acquaintances | If an GC algorithm does not support Inverse Acquaintances then the algorithm has to trace the entire reachability set of the unblocked actor through which an actor pass message to the root. | When to maintain the Inverse Acquaintances | Inverse acquaintances can be maintained at the time of garbage collection. |

Since Mailqueue objects and Inverse acquaintances are introduced in the transformed graph, exactly twice the number of objects and thrice the no of references are added as overhead.

## 3.2 Direct and Indirect Back Pointers by Wei Jen Wang

### 3.2.1 Analysis of Cost

To understand the impact of actor garbage collection, we measure actor garbage collection using four different mechanisms: NO-GC, GDP, LGC, and CDGC. By using these mechanisms, we can understand the overhead each actor garbage collection algorithm imposes on the actor system. The mechanisms are described as follows:

> No-GC: Data structures and algorithms for actor garbage collection are not used.
>
> GDP: The local garbage collector is not activated. Only the garbage detection protocol (the implementation of the pseudo-root approach) is used.
>
> LGC: The local garbage collector is activated every n seconds or in the case of insufficient memory (n=2 for the tests in this chapter).
>
> CDGC: The logically centralized garbage collector is activated every m seconds or in the case of insufficient memory (m=20 for the tests).

We developed three different benchmark applications using to measure the impact of our local actor garbage collection mechanism. These applications are Fibonacci number (Fib), N queens number (NQ), and Matrix multiplication (MX). Each application is executed on a dual-core processor Sun Blade 1000s machine, equipped with two 750 MHz processors and 2 GB of RAM. The operating system used was SunOS 5.10 and the Java VM was Java HotSpot Client VM (build 1.4.1). The applications are described as follows:

> Fibonacci number (Fib): Fibonacci number, abbreviated as Fib, takes one argument k and then computes the k-th Fibonacci number concurrently. It is a coordinated tree-structure computation. When k ≤ 30, the application sequentially computes the k-th Fibonacci number.

N queens number (NQ): N queens number, abbreviated as NQ, takes one argument to calculate the total solutions of the N queens problem by creating $(N-1) \times (N-2)$ actors for parallel execution and one actor for coordination.

Matrix multiplication (MX): Matrix multiplication, abbreviated as MX, requires two files for application arguments, each of which contains a matrix. The application calculates one matrix multiplication of the given two matrices.

We also developed four distributed benchmark applications. They are performed on four dual-core processor Sun Blade 1000s machines. The distributed benchmark applications are described as follows:

Distributed Fibonacci number with locality (Dfibl): Dfibl optimizes the number of inter-node messages by locating four sub-computing-trees at each computing node.

Distributed Fibonacci number without locality (Dfibn): Dfibn distributes the actors in a breadth-first-search manner.

Distributed N queens number (DNQ): DNQ equally distributes the actors to four computing nodes.

Distributed Matrix multiplication (DMX): DMX divides the first input matrix into four sub-matrices, sends the sub-matrices and the second matrix to four computing nodes, performs one matrix multiplication operation, and then merges the data at the computing node that initializes the computation.

The local experimental results are shown in Table 5, and the distributed results are in Table 6 Each result of a benchmark application is the average of ten execution times. Notice that Real represents the total real execution time to get the computing result, while CPU represents the total CPU time of both processors to get the computing result. CPU time can be bigger than Real time because the machine to test has two CPUs and CPU time is equal to the sum of the individual CPU time. The average GDP real time overhead of local experimental results is 20.5%; the average GDP CPU time overhead of local experimental results is 16%; the average LGC+GDP Real time overhead of local experimental results is 24%; the average LGC+GDP CPU time overhead of local experimental results is 19%; the average LGC+GDP+CDGC Real time overhead of experimental results is 19%;

Table 5: Local Experimental Results

| Application(Arguments) /Number of Actors | NO-GC (Real/CPU) | GDP (Real/CPU) | LGC+GDP (Real/CPU) | GDP OVERHEAD (Real/CPU) | LGC+GDP OVERHEAD (Real/CPU) |
|---|---|---|---|---|---|
| Fib(38)/109 | 1.70/2.57 | 2.14/3.07 | 2.13/3.08 | 25%/20% | 25%/20% |
| Fib(41)/465 | 5.09/8.66 | 6.20/10.21 | 6.63/10.54 | 22%/18% | 30%/22% |
| NQ(13)/133 | 1.84/2.16 | 2.42/2.55 | 2.55/2.79 | 32%/18% | 39%/29% |
| NQ(15)/183 | 6.72/11.58 | 7.63/12.84 | 7.97/13.30 | 14%/11% | 19%/15% |
| MX($100^2$)/3 | 1.84/1.93 | 2.16/2.24 | 2.16/2.24 | 17%/16% | 17%/16% |
| MX($150^2$)/3 | 2.63/2.84 | 2.97/3.17 | 3.03/3.20 | 13%/11% | 15%/13% |

Table 6: Distributed Results

| Application(Arguments)/Number of Actors | NO-GC | LGC+GDP+CDGC | LGC+GDP+CDGC OVERHEAD |
|---|---|---|---|
| Dfibl(39)/177 | 1.722 | 2.091 | 21% |
| Dfibl(42)/753 | 3.974 | 4.957 | 25% |
| Dfibn(39)/177 | 3.216 | 3.761 | 17% |
| Dfibn(42)/753 | 8.527 | 9.940 | 17% |
| DNQ(16)/211 | 13.120 | 17.531 | 34% |
| DNQ(18)/273 | 426.151 | 461.757 | 8% |
| DMX($100^2$)/5 | 6.165 | 6.715 | 9% |
| DMX($150^2$)/5 | 39.011 | 38.955 | 0% |

### 3.2.2 Analysis of Performance

Table 7 shows the various parameters which affects the performance together with the advantages, disadvantages and a possible suggestion to overcome them.

Table 7: Performance Analysis

| Parameter | Advantage | Disadvantage | Suggestion |
|---|---|---|---|
| Scanning the reference graph twice | Scans the reference graph only twice for marking, and has linear time complexity of O(V +E) and extra space complexity O(V +E). | Processing twice the graph is an Overhead | Use one extra marking variable in each actor, and scan the reference graph once |

## 4. Conclusion

In the transformation algorithm given by Abhay Vardhan Mailqueue objects and Inverse acquaintances are introduced in the transformed graph. Exactly twice the number of objects and thrice the number of references are added as overhead. Compared to Abhay Vardhan's transformation method Wei Jen Wang's method is efficient since there are no mail queue objects. The number of references is also less in Wei Jen Wang's method since inverse acquaintances are not added for all nodes. The back pointer algorithm requires scanning the reference graph twice which is again an overhead. Back Pointer algorithm has linear time complexity of O (V +E) and extra space complexity O (V +E). Only these two algorithms are available for transformation of active reference graph into passive reference graph. This area has to be further researched upon in the coming days to minimise the overheads caused by transformation.